\documentclass[twocolumn, natbib]{svjour3}

\usepackage{graphicx}
\usepackage{amsmath}
\usepackage{amssymb}
\usepackage{setspace}

\begin{document}
\title{Anisotropy in Fracking: A Percolation Model for Observed Microseismicity}

\author{J. Quinn Norris \and Donald L. Turcotte \and John B. Rundle}
\institute{J. Quinn Norris \and John B. Rundle \at Department of Physics, One Shields Ave., University of California, Davis, CA 95616, USA. \email{jqnorris@ucdavis.edu}
\and
Donald L. Turcotte \and John B. Rundle \at Department of Earth and Planetary Sciences, One Shields Ave., University of California, Davis, CA 95616, USA.
\and
John B. Rundle \at Santa Fe Institute, Santa Fe, NM 87501, USA.}

\titlerunning{Anisotropy in Fracking}

\maketitle

\begin{abstract}
 Hydraulic fracturing (fracking) using high pressures and a low viscosity fluid allow the extraction of large quantiles of oil and gas from very low permeability shale formations. The initial production of oil and gas at depth leads to high pressures and an extensive distribution of natural fractures which reduce the pressures. With time these fractures heal, sealing the remaining oil and gas in place. High volume fracking opens the healed fractures allowing the oil and gas to flow the horizontal productions wells. We model the injection process using invasion percolation. We utilize a 2D square lattice of bonds to model the sealed natural fractures. The bonds are assigned random strengths and the fluid, injected at a point, opens the weakest bond adjacent to the growing cluster of opened bonds. Our model exhibits burst dynamics in which the clusters extends rapidly into regions with weak bonds. We associate these bursts with the microseismic activity generated by fracking injections. A principal object of this paper is to study the role of anisotropic stress distributions. Bonds in the $y$-direction are assigned higher random strengths than bonds in the $x$-direction. We illustrate the spatial distribution of clusters and the spatial distribution of bursts (small earthquakes) for several degrees of anisotropy. The results are compared with observed distributions of microseismicity in a fracking injection. Both our bursts and the observed microseismicity satisfy Gutenberg-Richter frequency-size statistics.
\end{abstract}

\section{Introduction to Fracking}
Fracking, the common term for hydraulic fracturing, dates back to the late 1940's with the first commercial applications in 1949. The original process was a secondary recovery method designed to enhance production in reservoirs where primary recovery had decreased to the point where production was no longer economical. By injecting a high viscosity fluid at high pressures into the reservoir rock, one or two large fractures were created that extended from the borehole. Also injected were large quantities of sand which ``propped'' the generated fractures open and allowed oil and gas to flow through the fractures to the borehole. We refer to this process as low-volume or traditional fracking. Traditional fracking is applied to conventional reservoirs with high permeability, typically sandstone reservoirs. Because of the high permeability, the oil and gas can readily migrate to the generated fractures and flow to the borehole where it can be extracted. 

Two developments in the 1980's allowed fracking to extract oil and gas from tight shale reservoirs where the natural formation permeability is too low for economic extraction using traditional methods. The first development was horizontal drilling. Because many formations are relatively thin and lie nearly horizontally, a vertical well can only access a limited volume of reservoir rock. By turning the well bore horizontal, using directional drilling, a single well can access a much larger volume. This reduces the cost of well drilling making the extraction more economical. The second development was ``slickwater.'' In many reservoirs the low natural matrix permeability prevents the flow of oil and gas to a well. In these reservoirs, significant flow can only occur along fractures. By injecting a low viscosity fluid at high pressure, a distributed network of fractures is generated. These fractures increase the permeability in the rock surrounding the borehole allowing oil and gas to flow to the borehole. The combination of horizontal drilling with ``slickwater'' has changed the nature of fracking to a method that can be used to extract oil and gas from tight shales.

Shales are important source rocks for oil and gas \citep{Tourtelot1979, Arthur1994}. It is estimated that a large fraction of gas and oil has been formed in black shales during anoxic periods \citep{Ulmishek1990, Klemme1991, Trabucho-Alexandre2012}. As oil and gas develop in a shale, they generate pressures sufficient to fracture the rock \citep{Olson2009}. Typical shales have extensive fracture networks and joint sets. If a shale is relatively old, there is a greater chance that the natural fractures have been sealed by the deposition of silica or carbonates \citep{Gale2007a}. Fracking seems to be effective only in tight reservoirs where the natural fractures have been sealed \citep{King2012}, examples are the Barnett Shale in Texas and the Bakken Shale in North Dakota. Large quantities of natural gas are now being extracted from the Barnett Shale and large quantities of oil are being extracted from the Bakken Shale. Fracking appears to be ineffective in increasing the production from shales where the natural fractures are open, examples are the Antrim Shale in Michigan and the Monterey Shale in California. In both cases production of oil and gas continues to decrease. Fracture permeability of shales allow the migration of oil and gas into overlying strata which typically have a higher permeability. In the overlying strata, the oil and gas flow into structural or stratigraphic traps. These traps are the traditional reservoirs from which the majority of oil and gas recovery has occurred.

\begin{figure}
\centering
 \includegraphics[width=3.3in]{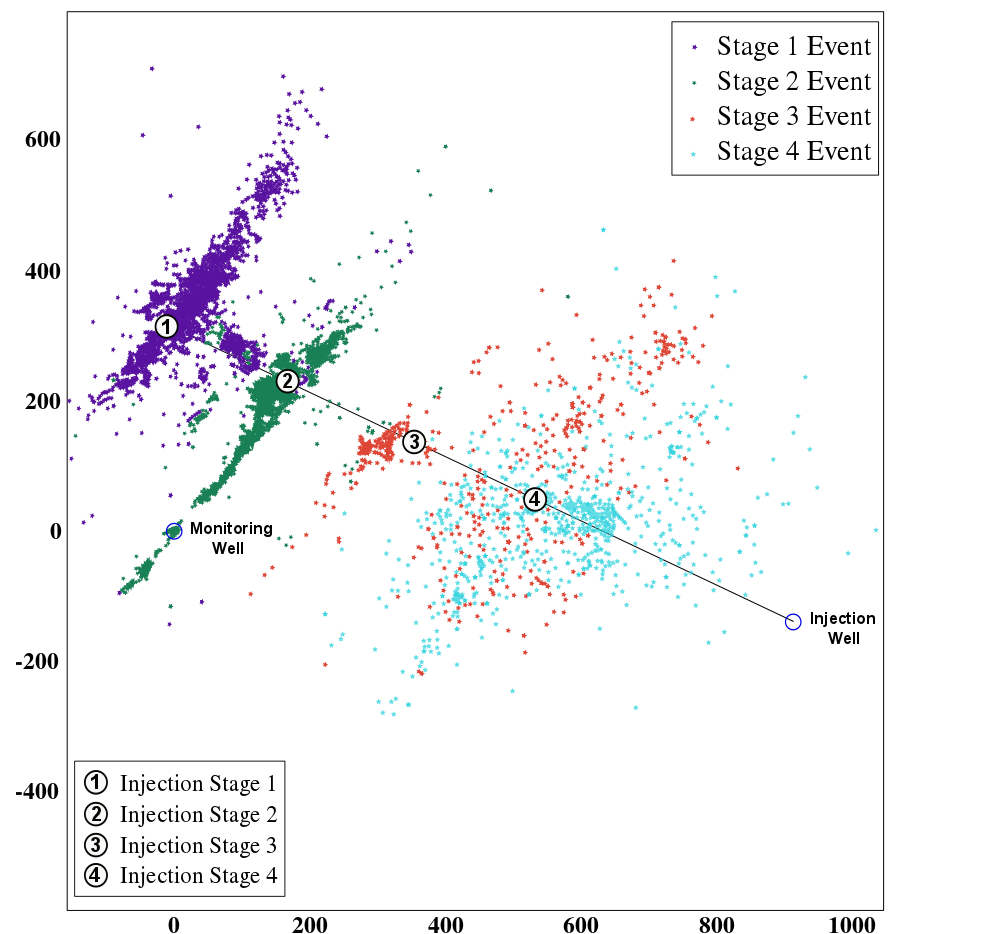}
 \caption{Map of the epiceneters of microseismicity associated with four fracks of the Barnett Shale \citep{Maxwell2011}. The colors correspond to the four injections and the axes are distances in meters from the monitoring well..}
 \label{fig:microseismicity}
\end{figure}

In order to understand the process and to optimize recovery, the fracking injections are often monitored using sensitive seismometers \citep{Warpinski2013}. In addition to recovery boreholes, one or two monitoring bore holes are often drilled. Sensitive seismometer arrays are placed along their lengths. The recorded microseismic data is used to determine the locations and magnitudes of the microseismic events that occur during a fracking treatment. This information can be used in real time to control the pressure, rate, and composition of the injected fluid. Because of the great depths, analysis of microseismic data is one of the primary methods used to understand the fracking process yet only 3\% of the fracking treatments performed in 2009 were microseismically monitored \citep{Zoback2010}. An example of microseismic data recorded during a four stage frack are shown in Figure \ref{fig:microseismicity}. Anisotropy plays a large role in fracking, with stress anisotropy being common. In a reservoir, the principle stresses are often not equal. Fractures grow perpendicular to the minimum principle stress and tend to be confined to the horizontal plane because the maximum principal stress in generally vertical. The distribution of microseismicity in Fig. \ref{fig:microseismicity} is clearly anisotropic.

As fracking has spread to more populated areas, such as the Marcellus Shale in Pennsylvania, there has been an increase in public concern over the safety of the process. The most common concerns are the potential to generate large earthquakes and the potential to contaminate drinking water. The largest recorded earthquake from a fracking treatment had a magnitude around three \citep{Ellsworth2013a}, much too small to cause any significant damage. However, the waste water generated during a fracking treatment is often re-injected into deep saline aquifers. There is increasing evidence that this re-injection causes larger earthquakes \citep{Keranen2013}. Additionally, because the microseismicity follows a power-law (Gutenberg-Richter) frequency-magnitude distribution \citep{Maxwell2011}, it is impossible to rule out the possibility of large events. It has been observed that drinking wells near fracking wells contain elevated levels of methane \citep{Osborn2011}. This contamination is not likely to be the result of the fracture network extending from the borehole at depth to the near-surface freshwater aquifer. Shale layers are typically three to five kilometers deep whereas freshwater aquifers are on the order of a hundred meters deep. Fractures running several kilometers would have seismic energy releases much greater that those observed during fracking treatments. These large fractures would also be undesirable from an engineering perspective. If the fracture network extends beyond the shale layer into the overlying strata, which often has a much lower permeability, significant leak-off can occur which reduces the effectiveness of the frack. The observed contamination is likely due to poor quality or damaged cement well casings.

Despite the lack of understanding of the processes and increased concern over the safety of fracking, there is very little publicly available research on fracking. This paper is part of an attempt to increase the availability of fracking research that can be used to understand the processes and risks associated with fracking.

\section{Introduction to Percolation}

Percolation theory has been used to study everything from conductivity \citep{Seager1974} to economics \citep{Cont2000}. Within geophysics, percolation theory has been used to study rock transport properties\citep{Gueguen1989}, earthquakes\citep{Otsuka1971, Sahimi1993}, and oil production \citep{King1999}. At its core, percolation is the study of connectivity. The classical random percolation problem is as follows: Given a random lattice of sites or bonds, what fraction of those bonds must be occupied for a cluster (a group of connected sites) to span from one side of a lattice to the other. This cluster is called the spanning cluster. The minimum occupation probability for which a spanning cluster exists on an infinite lattice (percolation threshold) has been shown to be a classical critical point. Near the percolation threshold, the statistics of the clusters are governed by power-laws analogous to the critical point of the liquid-gas phase transition. For a more complete introduction to percolation theory see \cite{Stauffer1994}

There have been many variants of this initial model \citep{Sahimi1994}. The variant closest to our model is called invasion percolation  \citep{Wilkinson1983} and has been used to study water flooding for oil recovery. Water flooding is a secondary recovery method where water is injected into a reservoir to drive oil and gas to the production well. In practice, several injection wells are drilled along one edge of an oil field and several production wells are drilled on the opposite edge. Water is injected from the injection wells and drives oil or gas to the production wells. In their model, the sites in a lattice are assigned random numbers on the interval [0,1). The sites along one edge are added to the perimeter of a growing cluster. The site on the perimeter with the smallest random number is invaded, and all sites adjacent to the invaded site are added to the perimeter. At each time step the site on the perimeter with the smallest random number is added to the single connected cluster. A later study involved injection from a single site with the cluster growing radially \citep{Wilkinson1984}. There are two major variants of the model, trapping and non-trapping, depending on whether the defending fluid is incompressible (trapping) or compressible (non-trapping) \citep{Knackstedt2002}. Non-trapping invasion percolation has been shown to belong to the same universality class as random percolation \citep{Knackstedt2002}. One of the properties of invasion percolation is that it displays self-organized criticality, i.e the dynamics take the system to a critical state. For a review of invasion percolation see \cite{Ebrahimi2010}.

Despite its relative simplicity, there are many aspects of percolation theory which are still unknown or not well studied. One of those aspects, addressed in this paper, is the role of anisotropy. Anisotropy is commonly introduced by occupying bonds in the horizontal direction with one probability $p_h$ and the bonds in the vertical direction with another probability $p_v$. The critical line for 2D bond percolation on a square lattice was determined by \cite{Sykes1963}. Since that time renormalization approaches have been used to explore anisotropic percolation both on and away from criticality \citep{Ikeda1979, Lobb1981, Kim1992}. The primary experimental work has been done in the field of material conductivity \citep{Smith1979, Mendelson1980, Balberg1983}. Initially, experimental results suggested that the introduction of anisotropy would cause changes to previously universal critical exponents \citep{Balberg1987}. More recent work suggests that anisotropic percolation should share universal isotropic exponents\citep{Han1991, Celzard2003}. The applicability of these results may be limited as our model provides a method for exploring the cross-over from one to two dimensional percolation. This cross-over requires a change in the critical exponents from their values in 2D. Similar cross-overs in percolation models have been studied previously \citep{Chame1984, Sotta2003}.

\cite{Herrmann1993a} studied anisotropic fracture propagation using a lattice of elastic beams subject to tensional failure. Fractal branching structures were obtained but the density of fractures were much lower than in percolation models

\section{Model}
Our model is an extension of the radial invasion model first studied by \cite{Wilkinson1984}. The isotropic version of our model has been given previously \citep{Norris2014}. The reservoir rock is assumed to have a network of natural fractures that have been sealed by deposition. We assume a point injection of a low viscosity fluid that breaks the seals as the fluid flows from the point of injection. We neglect the viscous pressure drop during flow and assume the fluid breaks the weakest seals as it flows through the matrix of preexisting fractures. The sealed fractures are represented by a lattice of bonds. Each bond is assumed to have a effective strength ($s$) which we represent with a random number. For simplicity we use a 2D square lattice of bonds.

Our justification for the applicability of the 2D-model is our interest in layered sedimentary deposits that have remained nearly horizontal. In many cases the target reservoir strata (the black shale) is relatively thin (say 100$\;$m) and the horizontal well is drilled within this strata. We hypothesize that fractures are confined to this target strata. The anisotropy we model is due to the anisotropy of the stress field in the layer. We assume the least principal stress is in the $y$-direction and the intermediate principal stress is in the $x$-direction. Thus the induced fractures will tend to propagate in the $x$-direction. Fractures tend to be oriented perpendicular to the direction of the least principal stress, the $y$-direction, and for this reason horizontal wells are drilled in this direction. 

In order to model the preferential fracture orientation due to the existing stress field in the rock, we assign random numbers (effective strengths) to bonds oriented in the $x$-direction on the interval $[0,1)$ and bonds oriented in the $y$-direction on the interval $[0,a)$ with $a>1$. When $a=1$ the model is isotropic. When $a=\infty$ the model becomes one-dimensional, with propagation only in the $x$-direction. Additionally, the tuning of $a$ provides a simple way of exploring the cross-over from 2D ($a=1$) to 1D ($a=0$) percolation.

The variable $a$ is a measure of the anisotropy and gives the relative likelihood of a fracture to propagate in a given direction. Thus in a simulation with $a=2$, the fracture network is twice as likely to grow in the $x$-direction than the $y$-direction. We relate this choice of anisotropy to the stresses in the rock.
\begin{equation}
 \sigma_{xx} = a \sigma_{yy}
\end{equation}
We can then interpret $a$ as being the ratio of the two principal horizontal stresses in the rock. 

\begin{figure}
\centering
 \includegraphics[width = 3.3in]{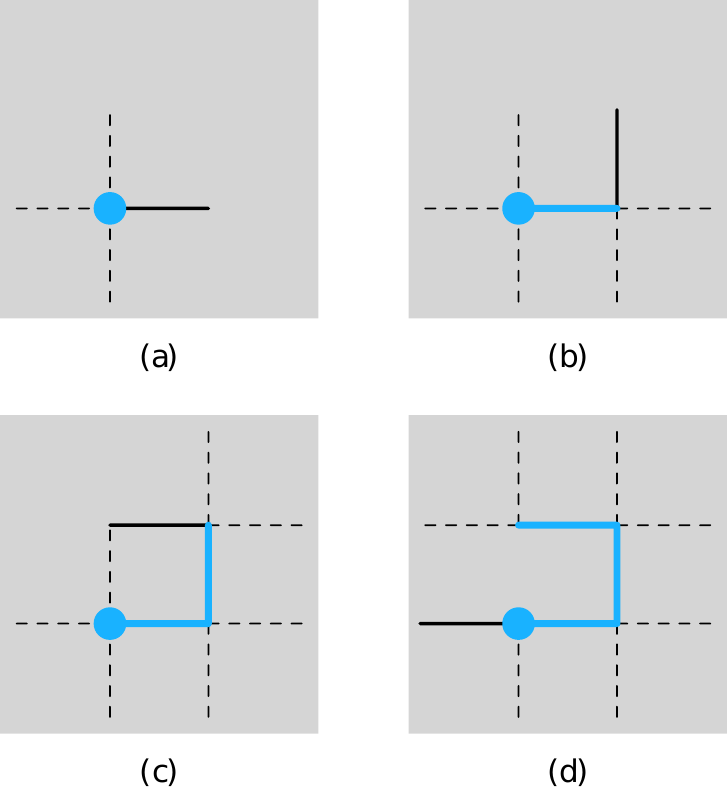}
 \caption{Illustration of our model. (a) Fluid is injected at the site shown. The four bonds to adjacent sites are also shown, the weakest bond (smallest $s$) is a solid line. (b) Three bonds to adjacent sites are added, the weakest of the six available bonds is a solid line. (c) Step b is repeated. (d) Step b is repeated, but the internal bond is removed.}
 \label{fig:model}
\end{figure}

Fluid is injected from a single site and the fracture network can grow in one of four directions as illustrated in Figure \ref{fig:model}a. These bonds are assigned anisotropic effective strengths as explained previously. The weakest bond (smallest $s$) fails and the fluid-filled fracture network grows in that direction as illustrated in Figure \ref{fig:model}b. The bond fails and fluid flows into the opened crack due to the pressure difference between the injected fluid and the surrounding rock. The new nearest neighbor bonds are assigned effective strengths and the process repeats. At each time step, the weakest bond on the perimeter of the growing fracture network fails, the fluid-filled fracture network grows in that direction, and new perimeter bonds are given effective strengths. Although the effective bond strength is not assigned until the bond joins the perimeter, the bond strength does not change once assigned (quenched disorder). If at anytime, the two ends of a bond belong to the growing fracture network as shown in Figure \ref{fig:model}c, the bond is removed from the simulation as shown in Figure \ref{fig:model}d. Because the differences in fluid pressure within the fluid filled fracture network are much smaller than the difference between the pressure in the surrounding rock, these bonds are much less likely to open and can be removed from the simulation. This bond removal step leads to a non-intersecting (loopless) fracture network. In our simulations we do not include an external boundary and the fracture network can grow indefinitely. Typically, we grow a cluster until a specified number of bonds have been added to the cluster. This can be thought of as limiting the volume of fluid injected during the fracking treatment. We refer to the number of bonds in a cluster as the mass ($M$) of the cluster.

\begin{figure}[]
\centering
 \includegraphics[]{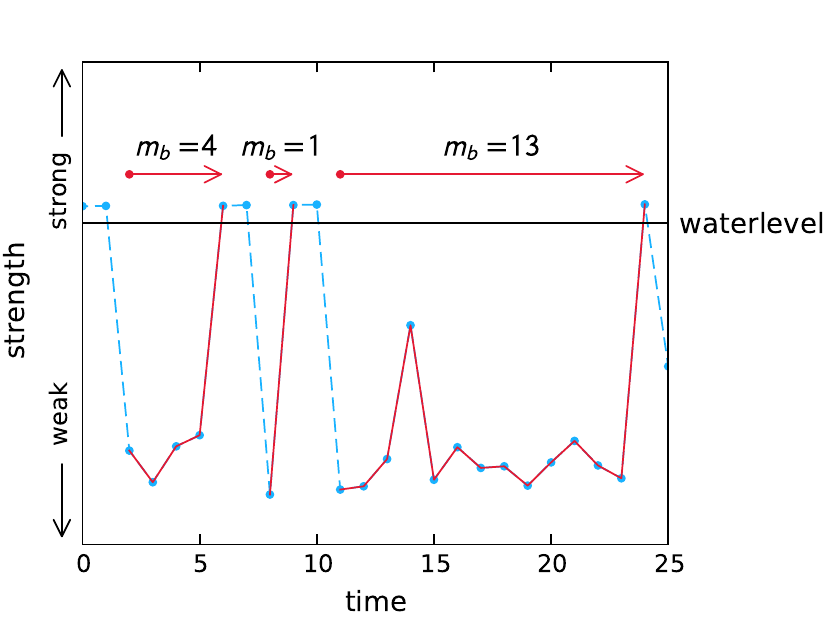}
 \caption{Illustration of our definition of a burst. A typical sequence of 25 opened bond strengths is given. A burst begins when an opened bond strength is below the waterlevel and ends when the strength is above the waterlevel. Three bursts with masses $m_b=4$, 1, and 13 are illustrated.}
 \label{fig:burst}
\end{figure}

In our and other invasion percolation models, growth occurs in bursts, the failure of a relatively strong bond is followed by the failure of a series of relatively weak bonds. We previously \citep{Norris2014} introduced a definition of a burst involving a waterlevel. A waterlevel is chosen. A burst begins when the strength of a failed bond falls below the chosen waterlevel. The burst continues until a failed bond's strength is greater than the chosen waterlevel. We refer to the number of bonds in a burst as the mass ($m_b$) of the burst. This definition is illustrated in Figure \ref{fig:burst}. By choosing a waterlevel just below the strength of the strongest failed bond in the fracture network, we obtain a power-law distribution of bursts \citep{Norris2014}. In our and other percolation models the strength of the strongest failed bond lies just below the percolation threshold of the lattice. This makes sense because in the absence of external constraints (stopping growth at a certain mass) the fracture network would grow to infinite size, percolating the infinite lattice. The power-law distribution of bursts lets us interpret bursts as the observed microseismic event generated during fracking treatments.

\section{Results}
We have performed simulations for several different values of anisotropy ($a$). These simulations take seconds on a desktop computer for even the largest runs making our model ideal for exploring parameter space. Our simulations are currently memory limited due to the large number of perimeter bonds that must be stored. One of the goals of this paper is to understand how the simulated microseismic events (bursts) are related to the underlying fracture network and the role anisotropy plays in the structure of both. We will first examine the statistics of grown clusters (simulated fracture networks) and then examine the statistics of bursts.

\subsection{Cluster Statistics}
In our model, the clusters represent the connected fracture network generated during a fracking treatment. It is important to understand the properties of this network to minimize risk and optimize production. For this paper, we are primarily interested in determining how anisotropy affects cluster properties.

\subsubsection{Images of Clusters}
\begin{figure*}[]
\centering
\includegraphics[]{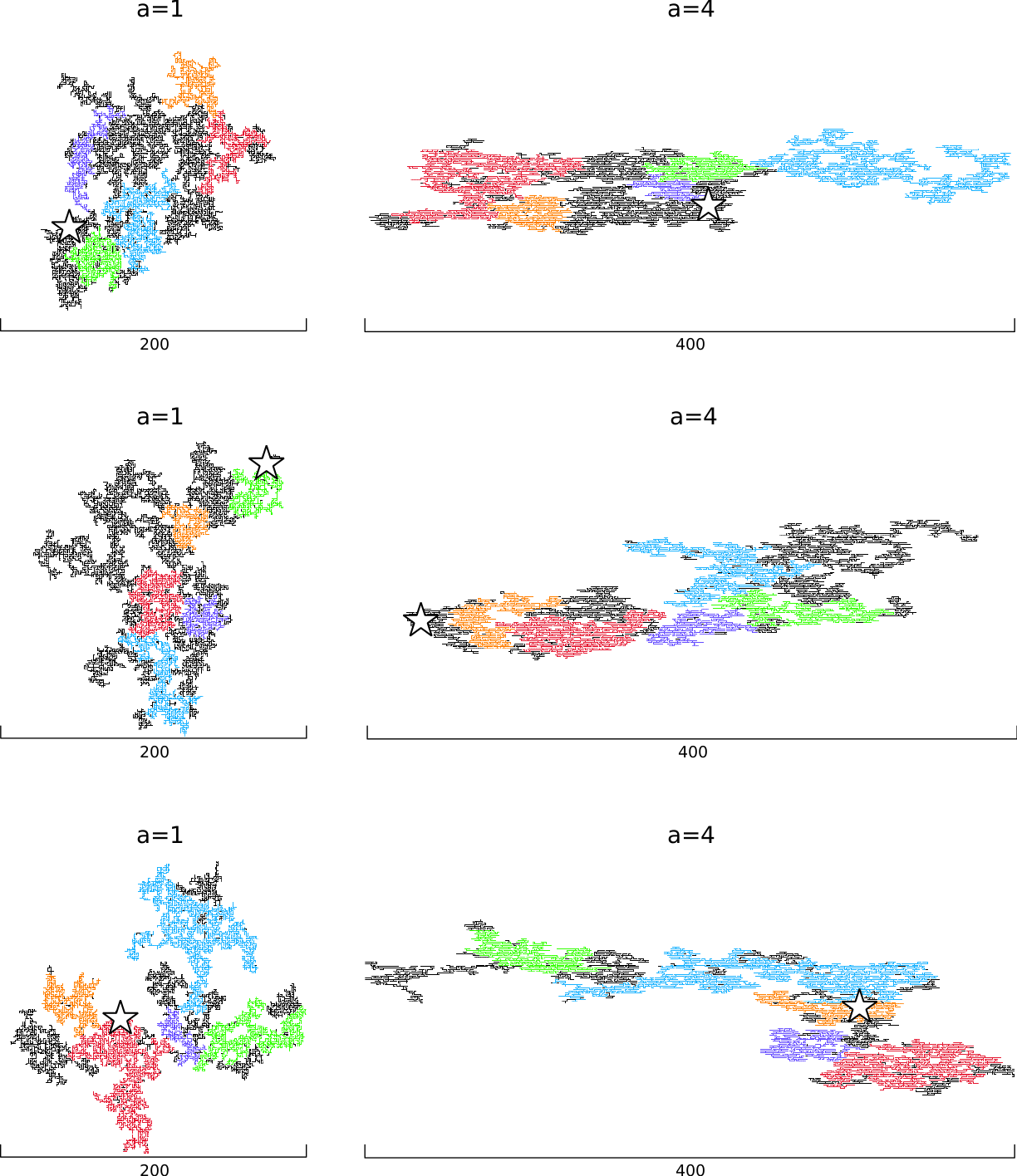}
\caption{Three examples of clusters with burst structures for $a=1$ and for $a=4$. In each realization $M=10,000$. The four largest bursts are shown in color. Smaller bursts and non-burst bonds are shown in black. The injection site for each cluster is shown as a star.}
\label{fig:images_of_clusters}
\end{figure*}

To get an idea of the general geometry of and variations between cluster realizations, we have generated three relatively small ($M=10,000$) clusters for two different anisotropies ($a=1$ and $a=4$. These clusters are shown in Figure \ref{fig:images_of_clusters}. The five largest bursts in each cluster have been colored, while the black bonds are smaller bursts and non-bursting bonds.

The clusters become more elongated with increasing anisotropy. This is expected as large anisotropies lead to weaker bonds in the $x$-direction. These weaker bonds provide the most likely paths for growth. As with other percolation models, there are many regions within a cluster that are completely surrounded by the cluster. The bonds on the boundaries of these cutoff regions are relatively strong with strengths greater than approximately $s=0.5$ and prevent further expansion of the cluster into the cutoff region. Bonds within the cutoff regions which are not on the boundaries can have any value of $s$, $0<s<1$. These cutoff regions are most easily observed in simulations where $a=1$. This is similar to the prevention of loops in self-avoiding random walks. 

Cluster growth is often asymmetrical about the point of injection despite a homogeneous distribution of strengths. In a few cases, growth occurs in nearly in a single direction. This shows that while the distribution may be symmetrical, a single realization may not exhibit that symmetry. It also shows that even small degrees of heterogeneity in an otherwise homogeneous material can result in large-scale inhomogeneous structures.

\subsubsection{Occupation Probability}
\begin{figure*}[]
\centering
\includegraphics[]{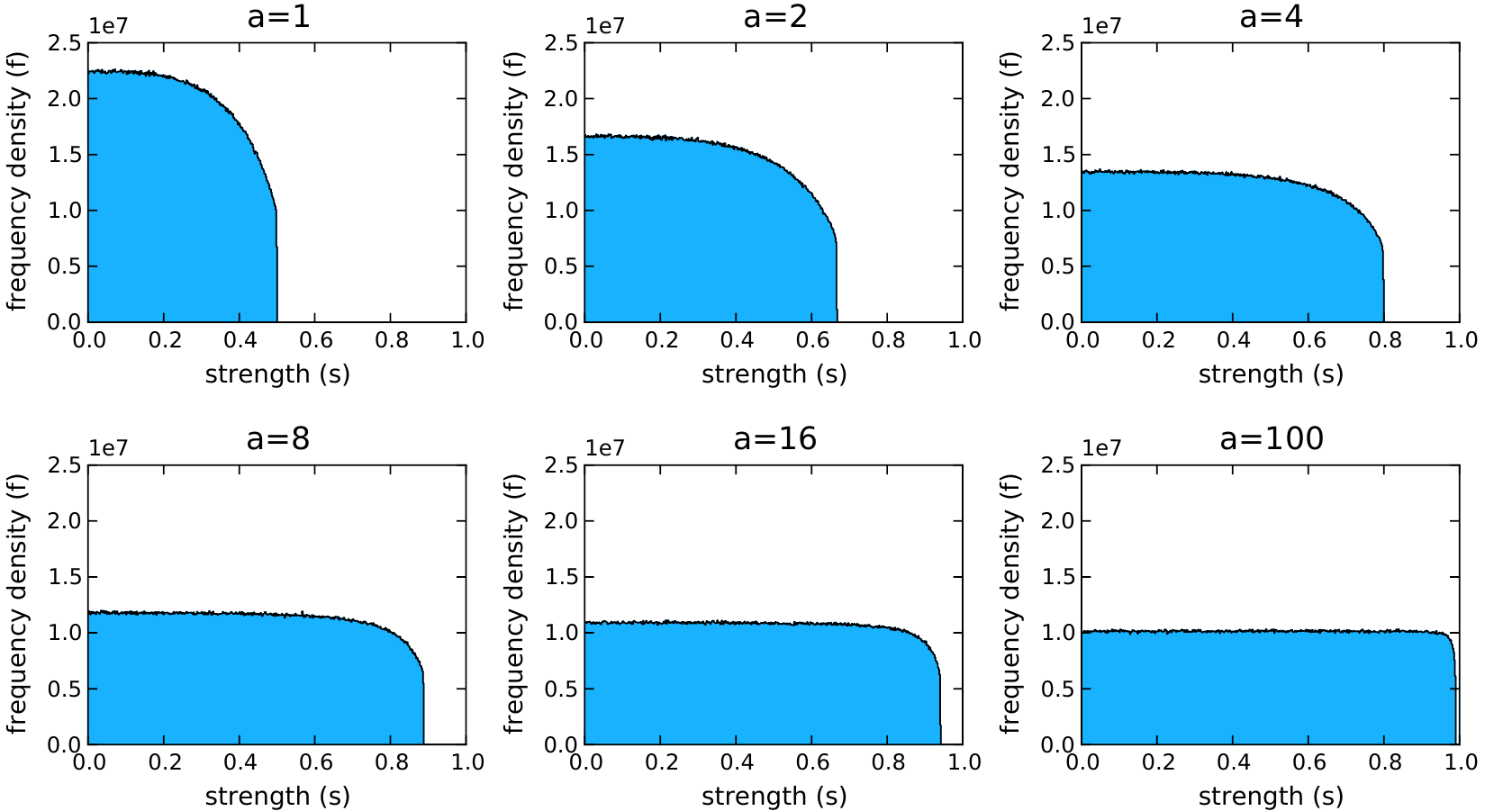}
\caption{Frequency densities of open bond strengths $f\left(s\right)$ are given as a function of bond strength $s$ for several values of $a$.}
\label{fig:bond_strength_frequency_density}
\end{figure*}

One property of interest is the distribution of bond strengths. In the isotropic case, we found that the frequency density ($f=\frac{\mathrm{d}N}{\mathrm{d}s}$) of bond strengths in the cluster shows a sharp cutoff near the critical point of the lattice \citep{Norris2014}. We have generated a single cluster of mass $M=10^7$ for six different anisotropies $a=1,2,4,8,16,100$ and calculated the frequency density of bond strengths as shown in Fig. \ref{fig:bond_strength_frequency_density}.

As in the isotropic case, the distributions of bond strengths for all the anisotropies considered show a sharp cutoff. The cutoff moves closer to 1 as anisotropy increases. Additionally as anisotropy increases, the distribution becomes more step-like. The critical line for 2D anisotropic bond percolation has been given previously from graph theory \citep{Sykes1963} and renormalization \citep{Arovas1983, Chaves1979}. The critical line in terms of $p_x$ and $p_y$, the occupation probability for bonds oriented in the $x$ and $y$ directions, is
\begin{equation}
 p_x + p_y = 1
 \label{eq:critical_line}
\end{equation}
In the isotropic case this equation gives $p_c = p_x = p_y = 0.5$. In our case, we do not have two different occupation probabilities, but a largest strength which in the isotropic case was equal to the critical occupation probability. To compare the largest strengths in the anisotropic case we rewrite Eq. \eqref{eq:critical_line} in terms of our anisotropic parameter $a$
\begin{equation}
 p + \frac{p}{a} = 1 \Rightarrow p =\frac{a}{a+1}
 \label{eq:critical_curve}
\end{equation}

\begin{figure}[]
\centering
 \includegraphics[]{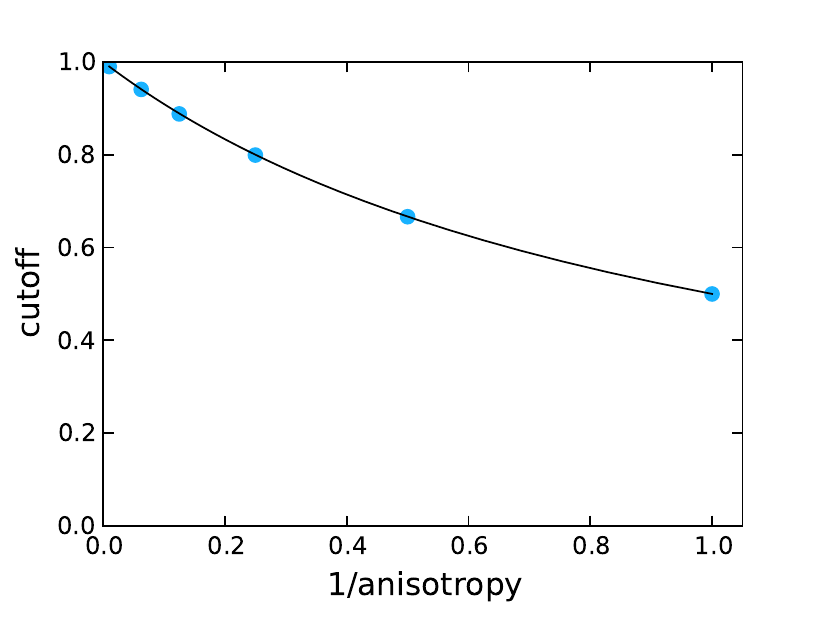}
 \caption{Bond strength cutoff as a function of 1/anisotropy ($\frac{1}{a}$). Solid circles are from the data given in Fig. \ref{fig:bond_strength_frequency_density} and the solid line is the critical point prediction from Eq. \eqref{eq:critical_curve}.}
 \label{fig:strength_cutoff}
\end{figure}

To determine whether the sharp cutoffs observed in Fig. \ref{fig:bond_strength_frequency_density} are near the critical value given in Eq. \eqref{eq:critical_line} we look for the largest strength in a cluster of mass $M=10^7$. Because small clusters often contain strengths above the cutoff, we determine the largest bond strength after an initial transient of 10 thousand bonds. The largest bond strength as a function of inverse anisotropy are shown in Fig. \ref{fig:strength_cutoff} along with the critical curve predicted by Eq. \ref{eq:critical_curve}. We find excellent agreement between the cutoffs and the critical values predicted by Eq. \ref{eq:critical_curve}, indicating that the largest bond strength is near the critical occupation probability for the lattice, even with the introduction of anisotropy.

\subsubsection{Fractal Dimension}
\begin{figure*}[]
\centering
 \includegraphics[]{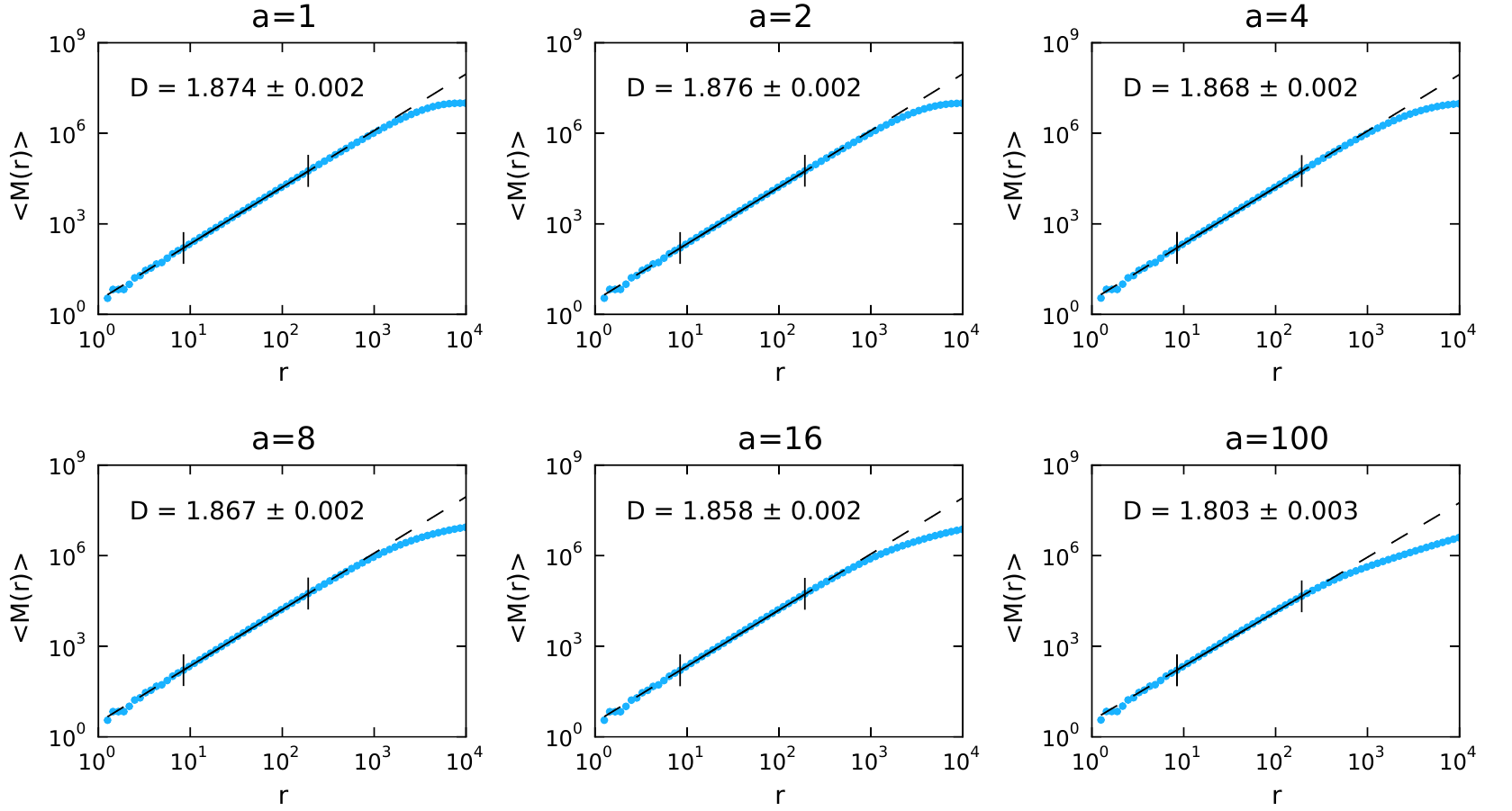}
 \caption{Dependence of the number of opened bonds $M$ contained within a circle of radius $r$ centered on the injection site on the radius $r$ for several values of the anisotropy parameter $a$. The best fits of Eq. \eqref{eq:fractal_dimension} to the data over the region between the vertical lines gives the fractal dimension $D$.}
 \label{fig:fractal_dimension}
\end{figure*}

One common measure used to distinguish between clusters of different types is the fractal dimension. To our knowledge the fractal dimension of anisotropic percolation clusters has not been measured previously. We follow the convention presented by \cite{Bunde2012} and define the fractal dimension ($D$) as the scaling exponent between the mass of the cluster $M$ and the radius from its center $r$
\begin{equation}
 M\left(r\right) \sim r^D
\end{equation}
Because we are interested in how much the reservoir is connected to the borehole, we measure the distance $r$ from the injection site (the borehole). In general, the borehole is not the location of the center of the mass of the cluster so different results may be obtained if distances are measured from the center of mass. To obtain good statistics, we generate 1000 clusters of mass $M=10^7$ for each value of anisotropy. For each cluster we center circles of varying radii ${r_1, r_2,...,r_i}$ on the injection site. For each circle we determine the mass of the cluster (number of bonds) contained within each circle ${M_1, M_2,...,M_i}$. We then take the logarithm of the radii and mass data and do a least-squares fit of aggregate log-log data to
\begin{equation}
\log{M} = D \log{r} + C
\label{eq:fractal_dimension}
\end{equation}
The average cluster mass as a function of radius along with the fit for several different anisotropies are shown on a log-log plot in Fig. \ref{fig:fractal_dimension}.
For large values of $r$ the cluster mass flattens out as entire clusters are contained within a circle of radius $r$ and the cluster begins to look more point-like. For small values of $r$ the discrete nature of the lattice causes variations in the masses. Because of these two factors we look for a linear region free both types of variation. The region used for the fit is shown in Fig. \ref{fig:fractal_dimension}. Because the region used is arbitrary, the uncertainty given is the uncertainty in the fit.

We see that the fractal dimension is weakly dependent on the anisotropy parameter. The fractal dimension only differs by $0.07$ between $a=1$ and $a=100$. Initially we thought that we could explore the cross-over to one-dimensional percolation; however, we see that anisotropies orders of magnitudes greater than those observed in reservoirs are required to significantly alter the fractal dimension.

\subsection{Burst Statistics}
Having quantified the anisotropic clusters generated by our model we not turn out attention to bursts. It is important to understand how the properties of the bursts are related to the properties of the underlying cluster. In fracking, this translates into understanding how the properties of the microseismic data are related to the underlying reactivated fracture network. Our definition of a burst requires the specification of a waterlevel just below the largest bond strength in the cluster. In this paper, we have shown that the largest bond strength in the cluster is near the critical occupation probability for the lattice. How close the largest bond strength is to the critical occupation probability depends on the size of the cluster. We have found that as clusters grow larger, the largest bond strength becomes asymptotically close to the critical occupation probability.

\begin{figure*}[]
\centering
\includegraphics[]{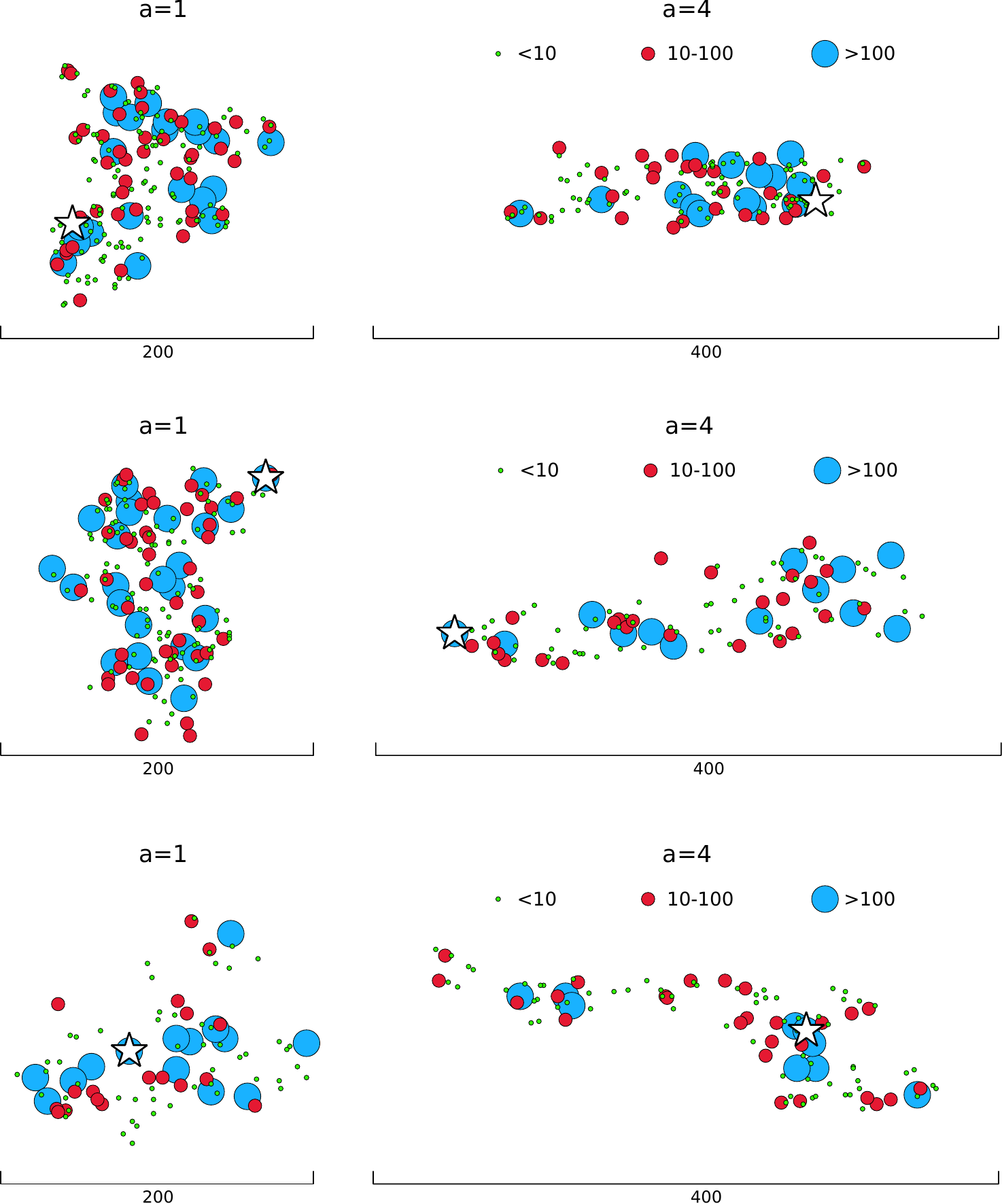}
 \caption{Epicenter locations of the bursts in the six realizations given in Fig. \ref{fig:images_of_clusters}. Different size circles correspond to different ranges of burst masses $m_b$.}
 \label{fig:burst_images}
\end{figure*}

For the six clusters shown in Fig. \ref{fig:images_of_clusters} we have determined the locations and sizes of the bursts using cutoffs of $0.47$ and $0.77$ for $a=1$ and $a=4$ respectively. We have plotted these bursts in Fig. \ref{fig:burst_images}. The burst markers are colored and scaled according to the size $m_b$ of the burst.
\subsubsection{Images of Bursts}

We see that the bursts and clusters cover roughly the same area and have a similar outline. This indicates that bursts are useful in determining the extent of cluster growth. This suggests that bursts could be used in real time to monitor the location and direction of cluster growth. We see that the bursts of different sizes seem to be more or less evenly distributed over the cluster, rather than larger bursts bunching in one location and small bursts bunching in another. The bursts shown are just a small sample and are meant to give a qualitative illustration of the spatial distribution of bursts.

To provide a more quantitative description of bursts we have determined the bursts for the same set of 1000 clusters of mass $M=10^7$ used to calculate the fractal dimension shown in Fig. \ref{fig:fractal_dimension}. Because these are relatively large clusters, a waterlevel very close to the critical occupation probability must be chosen. The critical occupation probabilities and chosen waterlevels for the six anisotropies considered are given in Table \ref{table:waterlevels}.
\begin{table}
\centering
\begin{tabular}{l c c}
Anisotropy & $p_c$ & Waterlevel \\
\hline
\\ [-1.5ex]
$a=1$ & 0.50000 & 0.49950 \\
$a=2$ & 0.66667 & 0.66610 \\
$a=4$ & 0.80000 & 0.79950 \\
$a=8$ & 0.88889 & 0.88830 \\
$a=16$ & 0.94118 & 0.94060 \\
$a=100$ & 0.99010 & 0.98995 \\
\end{tabular}
\caption{For the anisotropy values $a$ we consider, we give the critical probabilities from Eq. \eqref{eq:critical_curve} and our waterlevel values.}
\label{table:waterlevels}
\end{table}

\begin{figure*}[]
\centering
 \includegraphics[]{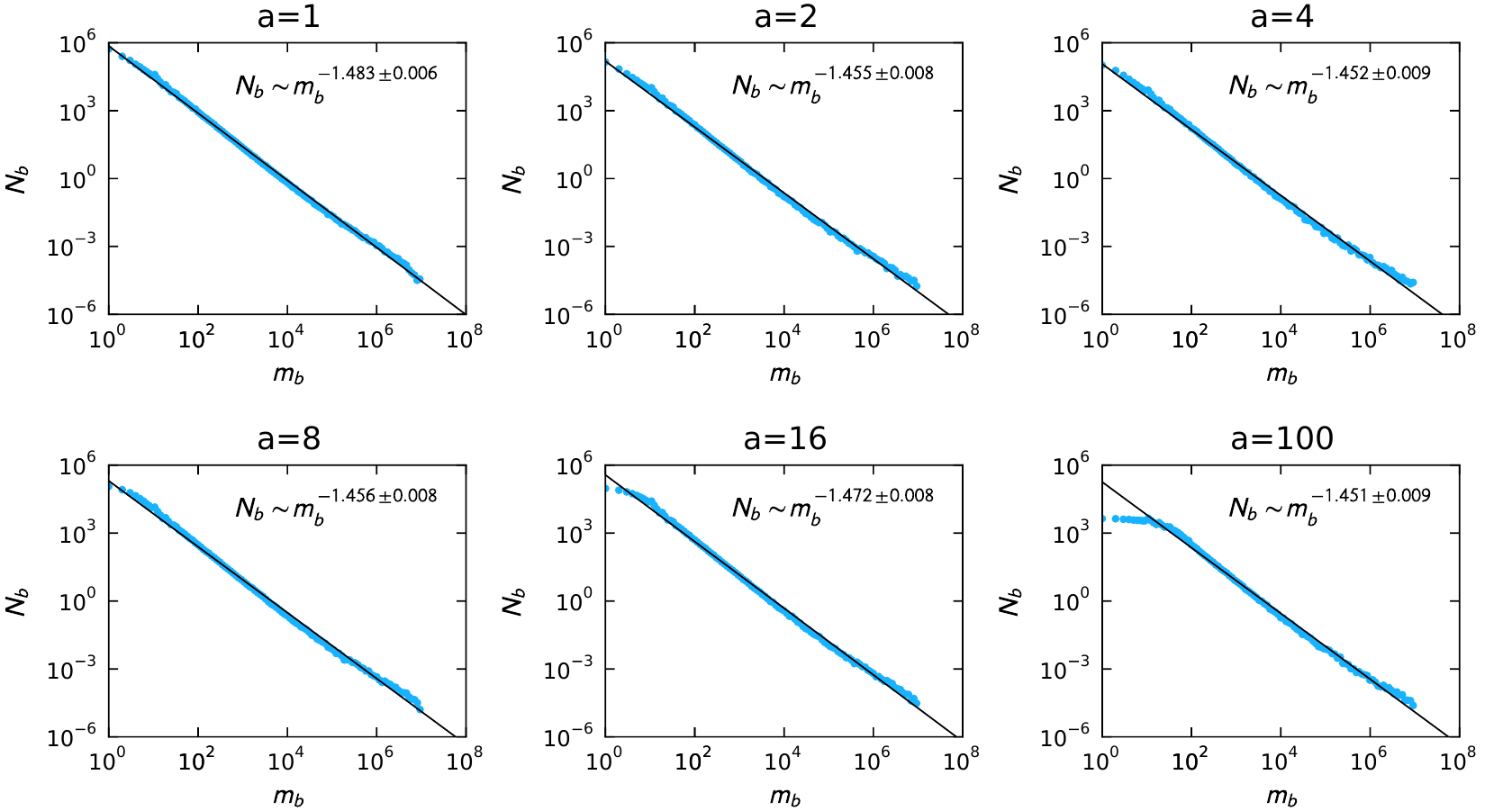}
 \caption{Dependence of the number of bursts $N_b$ with mass $m_b$ on $m_b$. These data are aggregated from 1000 runs with $M=10^7$ for each run. The best fit correlation with Eq. \eqref{eq:power_law} is shown for each value of $a$.}
 \label{fig:burst_size_distribution}
\end{figure*}

Using these values, we determine the non-cumulative burst frequency-size distribution for each realization. For each anisotropy considered, we aggregate the data. For larger bursts the aggregate data is sparse with zero or one bursts of a given mass. In this case, it is standard treatment to bin the data \citep{Malamud2004}. For each anisotropy considered, we bin the data and do a linear least-squares fit of the log-log data to the power-law distribution
\begin{equation}
 N_b \sim m_b^{-\tau}
 \label{eq:power_law}
\end{equation}
For larger anisotropies there is a rollover for small bursts. This region becomes significant for $a=16$ and $a=100$ and has been excluded from the fit. In Fig. \ref{fig:burst_size_distribution}, we give the binned data and fit for each value of anisotropy considered.
\subsubsection{Frequency-Size Distributions}

For each anisotropy value considered, we find excellent agreement with Eq. \eqref{eq:power_law}. The exponent takes values near $1.46$, suggesting that the slope is unchanged by the introduction of anisotropy. The rollover observed for $a=16$ and $a=100$ appears to be increasing with anisotropy. As the anisotropy becomes very large this rollover may dominate the distribution.

Because this is a non-cumulative distribution, the b-value for bursts generated by our model is $b=\tau-1$ and $b \approx .46$ for all anisotropies considered. This is significantly lower than the $b=2$ typically reported for microseismic data \cite{Maxwell2011}. More recent work \cite{Tafti2013} on injection into geothermal reservoirs has found $b \approx 1.3$. We note that while the b-value of our model is significantly lower than that observed during fracking treatments, our simulation is only in two-dimensions and moving to three dimensions might significantly change the b-value.

\subsubsection{Burst Fractal Dimension}
To obtain a comparison between clusters and bursts we calculate the fractal dimension for the bursts generated using our model. To obtain the fractal dimension we use the correlation function \citep{Hirata1987}
\begin{equation}
 C\left(r\right) = \frac{2}{N\left(N-1\right)}N_r\left(R < r\right),
 \label{eq:correlation}
\end{equation}
where $N$ is the total number of pairs of events and $N_r\left(R < r\right)$ is the number of pairs of events whose separation is less than $r$. If the burst distribution is fractal, the correlation should follow a power-law
\begin{equation}
 C\left(r\right) \sim r^D.
 \label{eq:correlation_dimension}
\end{equation}
The fractal dimension $D$ is sometimes called the correlation dimension and has been used previously to compare the fractal dimension of earthquakes to the fractal dimension of percolation clusters \cite{Tafti2013}. Using the same realizations as before, we calculate the correlation defined in Eq. \ref{eq:correlation} for the burst hypocenters. For each realization we fit the linear region to Eq. \eqref{eq:correlation_dimension}. If the realization does not have a large enough linear region or has two different linear realizations it is discarded. For the six anisotropies considered, we have calculated the mean and standard deviation of the fractal dimensions of the remaining realizations. These are given in Table \ref{table:fractal_dimension}.
\begin{table}
\centering
\begin{tabular}{l c}
Anisotropy & Burst Fractal Dimension (D) \\
\hline
\\ [-1.5ex]
$a=1$ & $1.84 \pm 0.11$ \\
$a=2$ & $1.82 \pm 0.15$ \\
$a=4$ & $1.81 \pm 0.17$ \\
$a=8$ & $1.79 \pm 0.16$ \\
$a=16$ & $1.77 \pm 0.17$ \\
$a=100$ & $1.49 \pm 0.25$ \\
\end{tabular}
\caption{The mean and standard deviations of the burst fractal dimensions $D$ are given for the anisotropy values $a$ that we consider.}
\label{table:fractal_dimension}
\end{table}

In all cases, the average fractal dimension of the bursts is less than the fractal dimension of the clusters. We also find that the difference increases with increasing anisotropy, However, for relatively small anisotropies the variation in fractal dimension is small and might not be significant in practical applications.

\subsubsection{Burst Anisotropy}
We also want to determine how the bursts are related to the underlying anisotropy in the bond strength distribution. As a simple measure of burst anisotropy, we calculated the standard deviations $\mathrm{s}_x$ and $\mathrm{s}_y$ in the $x$ and $y$ locations of the burst hypocenters relative to the injection point. This measure gives the aspect-ratio of the spatial distribution of bursts. Using the same 1000 realizations of $M=10^7$, we aggregated the burst hypocenters and computed the means and standard deviations. In performing the averages, we found that the results did not change significantly if the bursts were weighted by their size. For simplicity, the results presented here are not weighted by burst size. Additionally, we calculated the anisotropy using a common method introduced by \cite{Family1985}. This method utilizes the ratio of the eigenvalues ($\lambda_1$ and $\lambda_2$) of the gyration tensor and gives an anisotropy in the range 0 (completely anisotropic) to 1 (fully isotropic). In order to compare this measure of anisotropy to our results, we take the square root of the inverse of the anisotropy defined by \cite{Family1985}.

\begin{figure}
\centering
 \includegraphics[]{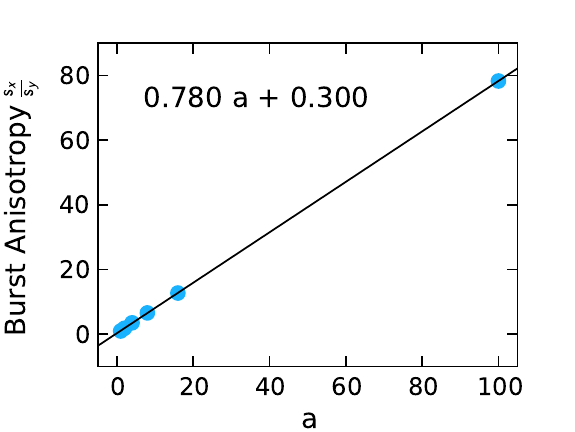}
 \caption{Dependence of the burst anisotropy $\frac{\mathrm{s}_x}{\mathrm{s}_y}$ on the bond strength anisotropy $a$. A linear correlation of the data is also shown.}
 \label{fig:anisotropies}
\end{figure}

For the simulations discussed above, we have determined the mean values of the ratio of standard deviations $\frac{\mathrm{s}_x}{\mathrm{s}_y}$ and the mean values of the square root of the ratio of eigenvalues $\sqrt{\frac{\lambda_2}{\lambda_1}}$. These results are given in Table \ref{table:anisotropies} for several values of the anisotropy parameter $a$. The two methods give similar values which are somewhat less than the anisotropy of the strength distribution $a$. In Fig. \ref{fig:anisotropies} we give the dependence of $\dfrac{\mathrm{s}_x}{\mathrm{s}_y}$ on $a$. We find a good fit to a linear dependence.
\begin{table}
\centering
\begin{tabular}{l c c c}
Anisotropy & $\dfrac{\mathrm{s}_x}{\mathrm{s}_y}$ & $\sqrt{\dfrac{\lambda_2}{\lambda_1}}$  \\
\hline
\\ [-1.5ex]
$a=1$ & 0.987 & 0.980 \\
$a=2$ & 1.843 & 1.851 \\
$a=4$ & 3.526 & 3.527 \\
$a=8$ & 6.589 & 6.589 \\
$a=16$ & 12.73 & 12.74 \\
$a=100$ & 78.29 & 78.94 \\
\end{tabular}
\caption{The mean values of the ratio of the standard deviations $\frac{\mathrm{s}_x}{\mathrm{s}_y}$ and the mean values of the square root of the ratio of eigenvalues $\sqrt{\frac{\lambda_2}{\lambda_1}}$ are given for several values of the anisotropy parameter $a$.}
\label{table:anisotropies}
\end{table}

\section{Discussion}
High volume fracking allows the extraction of oil and gas from tightly sealed shale reservoirs. During the initial formations of the reservoirs natural hydraulic fractures are generated by the high pressure associated with gas and oil generation. With time, these natural hydraulic fractures are sealed by chemical deposition. The injection of a high pressure, low viscosity fluid penetrates the formation reopening the sealed fractures. This allows the migration of oil and gas to the horizontal injection/production wells.

In order to model the injection process, we utilized invasion percolation. We assume a 2D square lattice of bonds. These bonds represent the preexisting array of natural fractures. Each bond is assigned a random strength and the weakest bond breaks at each time step. This represents the migration of the injected fluid through the sealed network of natural fractures. This migration occurs in bursts as the fluid enters a region of weak bonds. We associate these bursts with the microseismicity that occurs during fracking injections. A primary focus of this paper is on the role of anisotropic strengths on injection patterns.

The examples of microseismicity associated with four fracking injections into the Barnett Shale, illustrated in Fig. \ref{fig:microseismicity}, clearly show strong anisotropy. It is of interest to compare this microseismicity with the modeled microseismicity given in Fig. \ref{fig:burst_images}. The state 1 injection in Fig. \ref{fig:microseismicity} is similar to the $a=4$ injections illustrated in Fig. \ref{fig:burst_images}. Both the modeled and the observed microseismicity exhibit Gutenberg-Richter frequency magnitude statistics, however the b-values differ.

Our model certainly involves a number of serious approximations. Our model utilizes a two-dimensional square grid. Actual fluid injections are clearly three dimensional but seismic observations indicate the flow tends to be confined to a relatively narrow horizontal layer. The preexisting natural fractures that the injection reopens tend to have spacings in the range 0.1 to 1 meter, but are only approximated by a square grid.

Our model neglects the pressure drops associated with the fluid flow. This is probably a good approximation between ``bursts'' (microseismic events) but significant pressure drops may occur during a ``burst.'' Our model also assumes that the assigned bonds strengths are uncorrelated in space, some spacial correlations may be expected in actual reservoirs.

Despite the assumptions, the geometries of the invading cluster and associated modeled microseismicity are certainly qualitatively similar to the patterns of injection indicated by observed microseismicity. As discussed in our introduction, high volume fracking is successful only if the natural fractures are largely sealed. Unsealed fractures allow the inject fluid to flow through them without producing the distributed damage required for production. An interesting future extension of this model would include some open fractures prior to injection to quantify the problems associated with fluid leakage through these fractures.
\begin{acknowledgements}
The research of JQN and JBR has been supported by a grant from the US Department of Energy to the University of California, Davis \#DE-FG02-04ER15568
\end{acknowledgements}


\begin{thebibliography}{46}
\providecommand{\natexlab}[1]{#1}
\providecommand{\url}[1]{{#1}}
\providecommand{\urlprefix}{URL }
\expandafter\ifx\csname urlstyle\endcsname\relax
  \providecommand{\doi}[1]{DOI~\discretionary{}{}{}#1}\else
  \providecommand{\doi}{DOI~\discretionary{}{}{}\begingroup
  \urlstyle{rm}\Url}\fi
\providecommand{\eprint}[2][]{\url{#2}}

\bibitem[{Arovas et~al(1983)Arovas, Bhatt, and Shapiro}]{Arovas1983}
Arovas D, Bhatt RN, Shapiro B (1983) {Anisotropic bond percolation in two
  dimensions}. Phys Rev B 28(3):1433--1437, \doi{10.1103/PhysRevB.28.1433}

\bibitem[{Arthur and Sageman(1994)}]{Arthur1994}
Arthur MA, Sageman BB (1994) {Marine black shales: Depositional mechanisms and
  environments of ancient deposits}. Annual Review of Earth and Planetary
  Sciences 22(1):499--551, \doi{10.1146/annurev.ea.22.050194.002435}

\bibitem[{Balberg(1987)}]{Balberg1987}
Balberg I (1987) {Tunneling and nonuniversal conductivity in composite
  materials}. Phys Rev Lett 59(12):1305--1308,
  \doi{10.1103/PhysRevLett.59.1305}

\bibitem[{Balberg et~al(1983)Balberg, Binenbaum, and Bozowski}]{Balberg1983}
Balberg I, Binenbaum N, Bozowski S (1983) {Anisotropic percolation in carbon
  black-polyvinylchloride composites}. Solid State Communications
  47(12):989--992,

\bibitem[{Bunde and Havlin(2012)}]{Bunde2012}
Bunde A, Havlin S (2012) {Fractals and Disordered Systems, 2nd Ed.} Springer
  London

\bibitem[{Celzard and Mar\^{e}ch\'{e}(2003)}]{Celzard2003}
Celzard A, Mar\^{e}ch\'{e} JF (2003) {Non-universal conductivity critical
  exponents in anisotropic percolating media: a new interpretation}. Physica A:
  Statistical Mechanics and its Applications 317(3–4):305--312

\bibitem[{Chame et~al(1984)Chame, de~Queiroz, and dos Santos}]{Chame1984}
Chame A, de~Queiroz SLA, dos Santos RR (1984) {Dimensional crossover in
  directed percolation}. Journal of Physics A: Mathematical and General
  17(12):L657

\bibitem[{Chaves et~al(1979)Chaves, Oliveira, de~Queiroz, and
  Riera}]{Chaves1979}
Chaves CM, Oliveira PM, de~Queiroz SLA, Riera R (1979) {Remarks on the
  percolation problem in anisotropic systems}. Progress of Theoretical Physics
  62(6):1550--1555, \doi{10.1143/PTP.62.1550}

\bibitem[{Cont and Bouchaud(2000)}]{Cont2000}
Cont R, Bouchaud JP (2000) {Herd behavoir and aggregate fluctuations in
  financial markets}. Macroeconomic Dynamics 4(02):170--196

\bibitem[{Ebrahimi(2010)}]{Ebrahimi2010}
Ebrahimi F (2010) {Invasion percolation: A computational algorithm for complex
  phenomena}. Computing in Science \& Engineering 12(2):84--93,
  \doi{10.1109/MCSE.2010.42}

\bibitem[{Ellsworth(2013)}]{Ellsworth2013a}
Ellsworth WL (2013) {Injection-induced earthquakes}. Science 341(6142),
  \doi{10.1126/science.1225942}

\bibitem[{Family et~al(1985)Family, Vicsek, and Meakin}]{Family1985}
Family F, Vicsek T, Meakin P (1985) {Are Random Fractal Clusters Isotropic?}
  Phys Rev Lett 55(7):641--644, \doi{10.1103/PhysRevLett.55.641}

\bibitem[{Gale et~al(2007)Gale, Reed, and Holder}]{Gale2007a}
Gale JFW, Reed RM, Holder J (2007) {Natural fractures in the Barnett Shale and
  their importance for hydraulic fracture treatments}. AAPG Bulletin
  91(4):603--622, \doi{10.1306/11010606061}

\bibitem[{Gueguen and Dienes(1989)}]{Gueguen1989}
Gueguen Y, Dienes J (1989) {Transport properties of rocks from statistics and
  percolation}. Mathematical Geology 21(1):1--13
  
\bibitem[{Han et~al(1991)Han, Lee, and Lee}]{Han1991}
Han KH, Lee JO, Lee SI (1991) {Confirmation of the universal conductivity
  critical exponent in a two-dimensional anisotropic system}. Phys Rev B
  44(13):6791--6793, \doi{10.1103/PhysRevB.44.6791}

\bibitem[{Herrmann et~al(1993)Herrmann, Sahimi, and
  Tzschichholz}]{Herrmann1993a}
Herrmann HJ, Sahimi M, Tzschichholz F (1993) {Examples of fractals in soil
  mechanics}. Fractals 01(04):795--805, \doi{10.1142/S0218348X93000824}

\bibitem[{Hirata et~al(1987)Hirata, Satoh, and Ito}]{Hirata1987}
Hirata T, Satoh T, Ito K (1987) {Fractal structure of spatial distribution of
  microfracturing in rock}. Geophysical Journal of the Royal Astronomical
  Society 90(2):369--374, \doi{10.1111/j.1365-246X.1987.tb00732.x}

\bibitem[{Ikeda(1979)}]{Ikeda1979}
Ikeda H (1979) {Percolation in anisotropic systems: — Real-space
  renormalization group —}. Progress of Theoretical Physics 61(3):842--849,
  \doi{10.1143/PTP.61.842}

\bibitem[{Keranen et~al(2013)Keranen, Savage, Abers, and Cochran}]{Keranen2013}
Keranen KM, Savage HM, Abers GA, Cochran ES (2013) {Potentially induced
  earthquakes in Oklahoma, USA: Links between wastewater injection and the 2011
  Mw 5.7 earthquake sequence}. Geology 41(6):699--702, \doi{10.1130/G34045.1}

\bibitem[{Kim and Lee(1992)}]{Kim1992}
Kim CS, Lee MH (1992) {Monte Carlo renormalization group studies of anisotropic
  bond percolation}. International Journal of Modern Physics B
  06(09):1505--1515, \doi{10.1142/S0217979292000700}

\bibitem[{King(2012)}]{King2012}
King G (2012) {Hydraulic fracturing 101: What every fepresentative,
  environmentalist, fegulator, feporter, investor, university researcher,
  neighbor and engineer should know about estimating frac risk and improving
  frac performance in unconventional gas and oil wells}. In: SPE Hydraulic
  Fracturing Technology Conference, 6-8 February, The Woodlands, Texas, USA

\bibitem[{King et~al(1999)King, Jr., Buldyrev, Dokholyan, Lee, Havlin, and
  Stanley}]{King1999}
King PR, Jr JS, Buldyrev SV, Dokholyan N, Lee Y, Havlin S, Stanley H (1999)
  {Predicting oil recovery using percolation}. Physica A: Statistical Mechanics
  and its Applications 266(1–4):107--114

\bibitem[{Klemme and Ulmishek(1991)}]{Klemme1991}
Klemme HD, Ulmishek GF (1991) {Effective petroleum source rocks of the world;
  stratigraphic distribution and controlling depositional factors}. AAPG
  Bulletin 75(12):1809--1851

\bibitem[{Knackstedt et~al(2002)Knackstedt, Sahimi, and
  Sheppard}]{Knackstedt2002}
Knackstedt MA, Sahimi M, Sheppard AP (2002) {Nonuniversality of invasion
  percolation in two-dimensional systems}. Physical Review E 65(3):35,101,
  \doi{10.1103/PhysRevE.65.035101}

\bibitem[{Lobb et~al(1981)Lobb, Frank, and Tinkham}]{Lobb1981}
Lobb CJ, Frank DJ, Tinkham M (1981) {Percolative conduction in anisotropic
  media: A renormalization-group approach}. Phys Rev B 23(5):2262--2268,
  \doi{10.1103/PhysRevB.23.2262}

\bibitem[{Malamud et~al(2004)Malamud, Turcotte, Guzzetti, and
  Reichenbach}]{Malamud2004}
Malamud BD, Turcotte DL, Guzzetti F, Reichenbach P (2004) {Landslide
  inventories and their statistical properties}. Earth Surface Processes and
  Landforms 29(6):687--711, \doi{10.1002/esp.1064}
\bibitem[{Maxwell(2011)}]{Maxwell2011}
Maxwell S (2011) {Microseismic hydraulic fracture imaging: The path toward
  optimizing shale gas production}. The Leading Edge 30(3):340--346,
  \doi{10.1190/1.3567266}

\bibitem[{Mendelson and Karioris(1980)}]{Mendelson1980}
Mendelson KS, Karioris FG (1980) {Percolation in two-dimensional,
  macroscopically anisotropic systems}. Journal of Physics C: Solid State
  Physics 13(33):6197

\bibitem[{Norris et~al(2014)Norris, Turcotte, and Rundle}]{Norris2014}
Norris JQ, Turcotte DL, Rundle JB (2014) {Loopless nontrapping
  invasion-percolation model for fracking}. Phys Rev E 89(2):22,119,
  \doi{10.1103/PhysRevE.89.022119}

\bibitem[{Olson et~al(2009)Olson, Laubach, and Lander}]{Olson2009}
Olson JE, Laubach SE, Lander RH (2009) {Natural fracture characterization in
  tight gas sandstones: Integrating mechanics and diagenesis}. AAPG Bulletin
  93(11):1535--1549, \doi{10.1306/08110909100}

\bibitem[{Osborn et~al(2011)Osborn, Vengosh, Warner, and Jackson}]{Osborn2011}
Osborn SG, Vengosh A, Warner NR, Jackson RB (2011) {Methane contamination of
  drinking water accompanying gas-well drilling and hydraulic fracturing}.
  Proceedings of the National Academy of Sciences 108(20):8172--6,
  \doi{10.1073/pnas.1100682108}

\bibitem[{Otsuka(1971)}]{Otsuka1971}
Otsuka M (1971) {A simulation of earthquake occurrence part 1. A mechanical
  model}. Zisin (Journal of the Seismological Society of Japan 2nd ser)
  24(1):13--25

\bibitem[{Sahimi(1994)}]{Sahimi1994}
Sahimi M (1994) {Applications of percolation theory}. Taylor \& Francis,
  London; Bristol, PA

\bibitem[{Sahimi et~al(1993)Sahimi, Robertson, and Sammis}]{Sahimi1993}
Sahimi M, Robertson MC, Sammis CG (1993) {Fractal distribution of earthquake
  hypocenters and its relation to fault patterns and percolation}. Phys Rev
  Lett 70(14):2186--2189, \doi{10.1103/PhysRevLett.70.2186}

\bibitem[{Seager and Pike(1974)}]{Seager1974}
Seager CH, Pike GE (1974) {Percolation and conductivity: A computer study. II}.
  Phys Rev B 10(4):1435--1446, \doi{10.1103/PhysRevB.10.1435}

\bibitem[{Smith and Lobb(1979)}]{Smith1979}
Smith LN, Lobb CJ (1979) {Percolation in two-dimensional conductor-insulator
  networks with controllable anisotropy}. Phys Rev B 20(9):3653--3658,
  \doi{10.1103/PhysRevB.20.3653}

\bibitem[{Sotta and Long(2003)}]{Sotta2003}
Sotta P, Long D (2003) {The crossover from 2D to 3D percolation: Theory and
  numerical simulations}. The European Physical Journal E 11(4):375--388,
  \doi{10.1140/epje/i2002-10161-6}

\bibitem[{Stauffer and Aharony(1994)}]{Stauffer1994}
Stauffer D, Aharony A (1994) {Introduction to Percolation Theory, 2nd Ed.}
  Taylor \& Francis Group

\bibitem[{Sykes and Essam(1963)}]{Sykes1963}
Sykes MF, Essam JW (1963) {Some exact critical percolation probabilities for
  bond and site problems in two dimensions}. Phys Rev Lett 10(1):3--4,
  \doi{10.1103/PhysRevLett.10.3}

\bibitem[{Tafti et~al(2013)Tafti, Sahimi, Aminzadeh, and Sammis}]{Tafti2013}
Tafti TA, Sahimi M, Aminzadeh F, Sammis CG (2013) {Use of microseismicity for
  determining the structure of the fracture network of large-scale porous
  media}. Phys Rev E 87(3):32,152, \doi{10.1103/PhysRevE.87.032152}

\bibitem[{Tourtelot(1979)}]{Tourtelot1979}
Tourtelot HA (1979) {Black shale; its deposition and diagenesis}. Clays and
  Clay Minerals 27(5):313--321

\bibitem[{Trabucho-Alexandre et~al(2012)Trabucho-Alexandre, Hay, and
  de~Boer}]{Trabucho-Alexandre2012}
Trabucho-Alexandre J, Hay WW, de~Boer PL (2012) {Phanerozoic environments of
  black shale deposition and the Wilson Cycle}. Solid Earth 3(1):29--42,
  \doi{10.5194/se-3-29-2012}

\bibitem[{Ulmishek and Klemme(1990)}]{Ulmishek1990}
Ulmishek GF, Klemme HD (1990) {Depositional controls, distribution, and
  effectiveness of world's petroleum source rocks}. Tech. rep., USGS Bulletin
  1931

\bibitem[{Warpinski(2013)}]{Warpinski2013}
Warpinski NR (2013) {Understanding Hydraulic Fracture Growth, Effectiveness,
  and Safety Through Microseismic Monitoring}. In: Effective and Sustainable
  Hydraulic Fracturing

\bibitem[{Wilkinson and Barsony(1984)}]{Wilkinson1984}
Wilkinson D, Barsony M (1984) {Monte Carlo study of invasion percolation
  clusters in two and three dimensions}. Journal of Physics A: Mathematical and
  General 17(3):L129--L135, \doi{10.1088/0305-4470/17/3/007}

\bibitem[{Wilkinson and Willemsen(1983)}]{Wilkinson1983}
Wilkinson D, Willemsen JF (1983) {Invasion percolation: a new form of
  percolation theory}. Journal of Physics A-Mathematical and General
  16(14):3365--3376, \doi{10.1088/0305-4470/16/14/028}

\bibitem[{Zoback et~al(2010)Zoback, Kitasei, and Copithorne}]{Zoback2010}
Zoback M, Kitasei S, Copithorne B (2010) {Addressing the Environmental Risks
  from Shale Gas Development}. Tech. rep., Worldwatch Institute: Natural Gas
  and Sustainable Energy Initiative

\end{thebibliography}

\end{document}